# Selective excitation of multipolar spoof plasmons using orbital angular momentum of light

Takashi Arikawa[1†], Tomoki Hiraoka[1†], Shohei Morimoto[1†], François Blanchard[2], Shuntaro Tani[3], Tomoko Tanaka[3], Kyosuke Sakai[4], Hiroki Kitajima[4], Keiji Sasaki[4] and Koichiro Tanaka[1,3] *

[1] Kyoto University, Department of Physics, Kyoto, 606-8502, Japan

[2] École de technologie supérieure (ÉTS), Department of Electrical Engineering, Montréal, Québec, H3C 1K3, Canada

[3] Kyoto University, Institute for Integrated Cell-Material Sciences (iCeMS), Kyoto, 606-8501, Japan

[4] Research Institute for Electronic Science, Hokkaido University, Hokkaido, 001-0020, Japan

† These authors contributed equally to this work
* email: kochan@scphys.kyoto-u.ac.jp

**The nature of light-matter interaction is governed by the spatial-temporal structures of a light field and material wavefunctions. The emergence of the light beam with transverse phase vortex, or equivalently orbital angular momentum (OAM) has been providing intriguing possibilities to induce unconventional optical transitions beyond the framework of the electric dipole interaction. The uniqueness stems from the OAM transfer from light to material, as demonstrated using the bound electron of a single trapped ion. However, many aspects of the vortex light-matter interaction are still unexplored especially in solids with extended electronic states. Here, we unambiguously visualized dipole-forbidden multipolar excitations in a solid-state electron system; spoof localized surface plasmon, selectively induced by the terahertz vortex beam. The results obey the selection rules governed by the conservation of the total angular momentum, which is numerically confirmed by the electromagnetic field analysis. Our results show light's OAM can be efficiently transferred to an elementary excitation in solids.**

The Fermi's golden rule defines whether an optical transition between quantum mechanical states is allowed or not. In the common case of the plane wave with no transverse phase structure in the long-wavelength regime (field gradient along the propagation direction is ignored), one obtains dipole selection rules, which has been successful in describing a variety of linear optical phenomena[1]. Several efforts have been made to break the selection rules and access dark electronic states by violating the uniform phase condition using oblique irradiation or steep field gradient in a metallic nanogap[2,3]. These methods simultaneously excite many modes due to the low symmetry, which is preferred for light energy harvesting. To selectively excite one particular dark mode, nonlinear optical techniques such as two-photon absorption and stimulated Raman scattering have been employed provided that strong light sources are available[4].

The vortex beam with the orbital angular momentum (OAM)[5] is a new and ideal tool for selectively exciting dipole-forbidden states by linear optical absorption at normal incidence. Its transverse phase is structured so that different selection rules from dipole ones are derived[6,7]. The selectivity can also be understood in terms of the OAM transfer from light to material. This is clearly demonstrated by the observation of the dipole-forbidden *S* to *D* state transition in the bound electron of a trapped ion, where the electron absorbs both spin angular momentum (SAM) and OAM from the vortex beam[8]. However, the transition probability of such a transition is very small due to the significant size mismatch between the diffraction-limited spot size of the vortex beam (~ 1000 nm) and the spatial extent of the electronic wavefunction (< tens of nm)[7,9].

In this context, electrons in solids with extended wavefunctions are expected to provide a promising platform for studying vortex light-matter interaction[10]. In fact, electromagnetic field analyses predict efficient OAM transfer from vortex beams to localized surface plasmons (LSPs) in a metallic disk of similar size to the diffraction-limited excitation beam[11]. In the simulations, multipolar modes with larger angular momentums (quadrupole, hexapole, ...) are selectively excited as a result of the OAM transfer, while Gaussian beam (plane wave) only splits positive and negative charges in an opposite direction, resulting in the dipole mode. In this study, we experimentally demonstrated this selective excitation using the spoof LSP; a low-frequency analogue of LSP, that exists around the surface of a periodically textured metallic disk[12-14]. It mimics LSP with tailor-made resonance frequencies depending on the geometric parameters of the corrugation. By properly designing the structures, we brought the resonance frequencies down to the terahertz (THz) frequency region, where nondestructive imaging of the near field distribution is possible in a time-resolved manner[15-17]. This allows us to unambiguously visualize characteristic near-field patterns around the corrugated disk and identify spoof LSP modes excited in the sample.

**Generation and detection of THz vortex beam**

Let us first characterize the THz excitation beam, which is measured without metallic structures. Linearly polarized coherent THz pulses are focused to the electro-optic (EO) crystal (LiNbO$_3$) by two lenses (Fig. 1a,b). When performing vortex beam experiments, we put a spiral phase plate (SPP) in the collimated part of the THz beam propagating in the *z*-direction[18-21]. We made two SPPs to generate vortex beams with the *z*-component of the OAM of ±$\hbar$ or ±2$\hbar$ centered at around 0.5 THz (SPP-1 and -2. See Methods and Supplementary information 2). Two-dimensional EO imaging was performed using a probe beam with a large spot size and a CMOS camera with a polarization analyzer unit (see Supplementary information 1). Figure 1d shows the snapshots of the THz pulse passing through the EO crystal taken without a SPP. A uniform phase (colour) of the electric field in the center suggests the original THz pulse is very close to the ideal Gaussian beam. Figures 1e and 1f show frequency-domain intensity and phase images at 0.5 THz, which also exhibit Gaussian beam nature; single intensity peak and flat phase distribution around the center. The snapshots shown in Fig. 1g was measured with the SPP-1. A vortex phase structure making a clockwise turn is clearly seen. The frequency-domain intensity (Fig. 1h) and phase (Fig. 1i) images at 0.5 THz show the characteristic features of the vortex beam with the OAM of +$\hbar$, i.e., donut-shape intensity profile and the 2π phase rotation with a center singular point (white cross). Similar experiments and analysis were done with the SPP-2 to generate the vortex beam with the *z*-component of the OAM of -2$\hbar$ at 0.5 THz. We can see a vortex rotating counter-clockwise (Fig. 1j) in the time-domain, and 4π phase rotation (Fig. 1l) with a center intensity null (Fig. 1k) in the frequency-domain. The non-ideal intensity profile suggests low OAM purity as characterized in the Supplementary information 2.

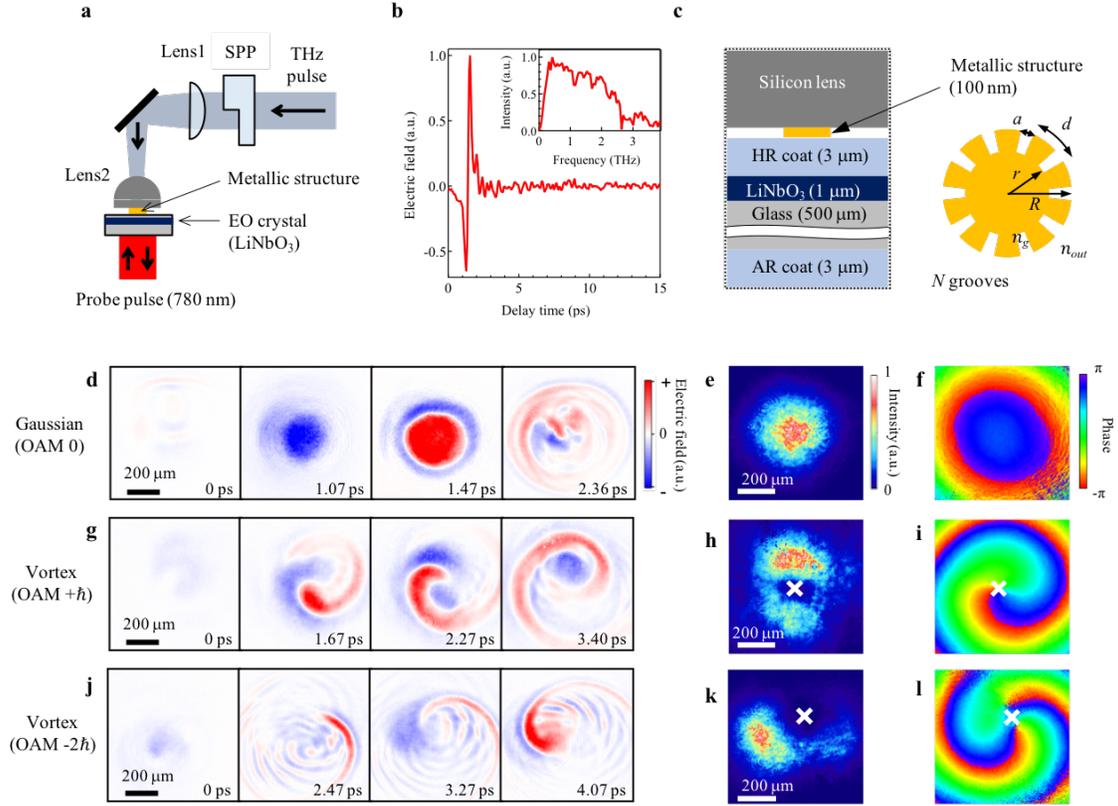

**Figure 1 | THz Gaussian and vortex beams. a.** Schematic of the experimental set-up. The SPP converts Gaussian beam into vortex beam. Two lenses (Lens1: Tsurupica lens, f = 50 mm. Lens2: 3-mm-thick, 2-mm-radius silicon bullet lens) were used to match the spot size of the excitation beam to the size of the metallic structure for efficient excitation. For details, see Supplementary information 1. **b.** Electric field waveform of the incident Gaussian THz pulse. The inset shows its frequency spectrum. **c.** Magnified view around the EO crystal (side view). 1-µm-thick, x-cut LiNbO$_3$ is mounted on a glass substrate. To perform EO sampling in the reflection geometry, the top and bottom surfaces of this substrate are coated with high-reflection (HR) and anti-reflection (AR) films (3 µm thick) for the probe pulse at 780 nm, respectively. The top view of the corrugated disk is schematically shown with the following structural parameters; inner radius (*r*), outer radius (*R*), periodicity (*d*), groove width (*a*), number of grooves (*N*). The refractive indices inside the groove and outside the disk are given by $n_g$ and $n_{out}$, respectively. **d,g,j.** Time-domain EO images of the Gaussian and vortex beams. **e,h,k.** Frequency-domain intensity images of the Gaussian and vortex beams at 0.5 THz. The intensity images are normalized by the maximum values. **f,i,l.** Phase images of the Gaussian and vortex beams at 0.5 THz.

**Direct observation of multipole spoof LSPs**

To visualize the near-field pattern due to spoof LSPs, we fabricated corrugated gold disks directly on the top surface of the EO substrate (Fig. 1c) using photolithographic and vacuum deposition technique. This allows us to sample electric field only a few μm (<< wavelength of a few hundred μm) away from the metallic structure[15-17]. Figure 2a shows five snapshots of the THz electric field around the sample excited by a linearly polarized Gaussian beam. Even after the incident THz pulse has passed through the sample, an electric field oscillation localized around the outer circle of the sample is observed (3rd to 5th frame), which indicates a resonant excitation of a spoof LSP. It has four nodes at fixed points along the outer circle. This is shown by the red curve in Fig. 2b, where we plot the electric field taken along the outer circle of the sample at 5.67 ps (4th frame) as a function of the azimuthal angle $\varphi$. Notice that the selected snapshots (3rd to 5th frame) display almost one oscillation period. These results show that a standing wave mode with a period of 4.2 ps (0.24 THz) is excited. The red curve in Fig. 2b nicely follows $\cos(2\varphi)$ drawn by the dashed black curve. This represents the expected electric field pattern when the dipole mode is excited in the sample, which is obtained by projecting the quasi-static electric field around the dipole onto the detection polarization axis ($\vec{e}_0$) in the experiment (see Methods for details). This confirms that the dipole mode is excited by the Gaussian beam.

Figure 2c shows selected EO images of the sample excited by the linearly polarized vortex beam with the OAM $+\hbar$. A localized electric field with six zero-crossing points along the outer circle is seen after the excitation. An azimuthal plot along the outer circle at 5.47 ps (3rd frame) agrees excellently with $\cos(3\varphi)$ curve, which is the expected functional form when the quadrupole mode exists (Fig. 2d). In contrast to the Gaussian beam excitation, we can see this electric field pattern travels along the outer circle in a clockwise direction. A similar experiment and analysis were carried out for the vortex beam excitation with the OAM $-2\hbar$ (Fig. 2e,f). Although not clear as above, multipolar field pattern with eight zero-crossing points can be recognized (Fig. 2e), which partly follows $\cos(4\varphi)$ curve suggesting the excitation of the hexapole mode (Fig. 2f).

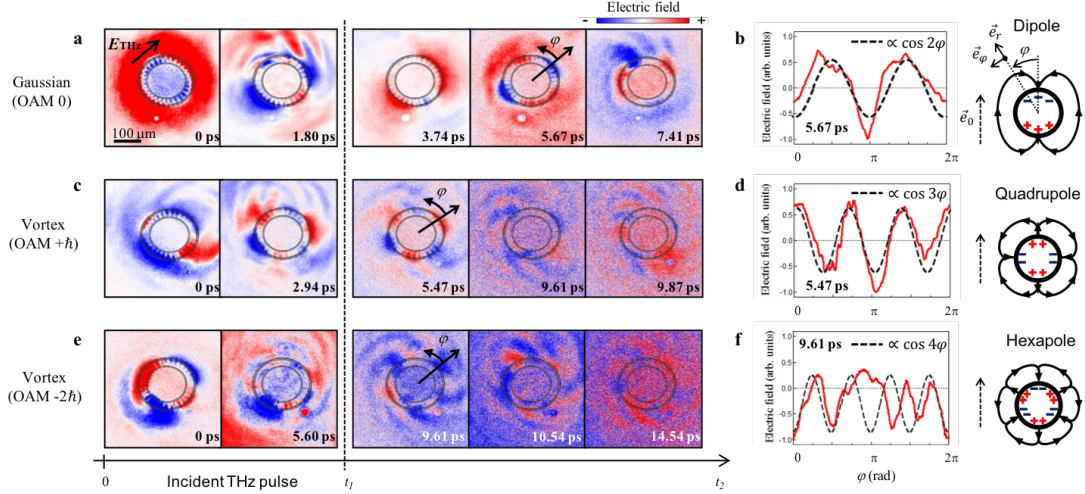

**Figure 2 | Selective excitation of multipole spoof LSPs.** Selected snapshots of the near-field evolution around the sample excited by **a.** Gaussian beam, **c.** vortex beam (OAM $+\hbar$), and **e.** vortex beam (OAM $-2\hbar$). The Double circle represents the position of the sample (inner radius $r$, outer radius $R$). The sample dimensions are $r = 70$ μm, $R = 100$ μm, $N = 30$, $a/d = 0.4$. The time origin (0 ps) is the time when the first positive peak of the incident pulse comes. The colour scales are optimized at each frame for the sake of clarity. **b,d,f.** The electric field taken along the outer circle of the sample as a function of the azimuthal angle $\varphi$ (red curves). The dashed cosine curves are expected electric field patterns when the modes depicted on the right are excited. The solid arrows schematically represent the quasi-static electric field around each mode. The cosine functions are obtained by projecting the quasi-static field onto the polarization axis ($\vec{e}_0$, dashed up arrow) detected in the experiment. $\vec{e}_r$ and $\vec{e}_\varphi$ are cylindrical unit vectors.

**Field pattern analysis**

These observations indicate that multipole spoof LSPs are excited by using vortex beams. To clearly illustrate this point, we performed the following analysis. The orthogonal functional forms of the electric field [$\cos(m\varphi)$, $m$: integer] for different LSP modes allow us to decompose the complicated field pattern into the superposition of each mode. See Methods for the proof of the one-to-one correspondence between the LSP modes and cosine functions for their electric field distributions. For this purpose, we focused on the electric field along the outer circle of the sample as we have done in Fig. 2b,d,f, but this time including the time dependence [$E(\varphi, t)$]. We expanded $E(\varphi, t)$ along both axes by two-dimensional Fourier transform to obtain $E(m, f) = \int_0^{2\pi} d\varphi \int_{t_1}^{t_2} dt E(\varphi, t) \cos(m\varphi - 2\pi f t)$. $E(m, f)$ then represents the frequency ($f$) spectrum of each LSP mode ($m = \pm 2$; dipole, $m = \pm 3$; quadrupole, and so on) excited in the sample. The positive (negative) sign of $m$ corresponds to the traveling wave mode in the clockwise (counter-clockwise) direction. To exclusively analyze the electric field generated by spoof LSPs, we only used data after the incident THz pulses have passed through the sample ($t_1$ to $t_2$ in Fig. 2). This enables us to minimize spurious signals due to the electric field pattern of the excitation pulse (see Supplementary information 2).

Figure 3a shows the frequency spectra of the dipole [$E(\pm 2, f)$], quadrupole [$E(\pm 3, f)$], and hexapole mode [$E(\pm 4, f)$] in the case of the Gaussian beam excitation. We can see only the dipole modes are excited at 0.24 THz. Note that both traveling modes with clockwise and counter-clockwise directions are equally excited, which is consistent with the standing wave observed in Fig. 2a. In contrast, the frequency spectra in the case of the vortex beam (+ℏ) excitation clearly show a peak in the quadrupole mode (Fig. 3b, middle). Furthermore, the positive index mode (red) is selectively excited as inferred from the clockwise traveling wave observed in Fig. 2c. The same analysis for the case of the vortex beam (-2ℏ) excitation clearly demonstrates the selective excitation of the negative hexapole mode (Fig. 3c, top). The selection of the inverse rotation modes is also confirmed by inverting the sign of the $z$-component of the OAM (See Supplementary Fig. S3). These results clearly demonstrate the selective and efficient excitation of the multipolar modes using the OAM of light. The excitation of the dipole mode in Fig. 3b,c and the negative quadrupole mode in Fig. 3c are explained by the residual Gaussian and vortex (-$\hbar$) beam components in the excitation pulse, respectively (see Supplementary Fig. S2). Although the simultaneous excitation of those modes complicates the electric field pattern in the time-domain (Fig. 2c,e), this analysis allows us to identify all spoof LSP modes excited in the sample.

This analysis also reveals the resonance frequencies of each spoof LSP mode and allows us to draw its dispersion relation (solid circles in Fig. 3d). As the number of poles increases, the resonance frequency becomes higher and shows a signature of saturation. This is well reproduced by the theoretical curve[22,23], which supports that multipole spoof LSPs are excited (see Methods for the detailed procedure of the fitting). The control of the resonance frequency of the spoof LSP by changing the dimensions of the corrugation is demonstrated in Supplementary Information 3.

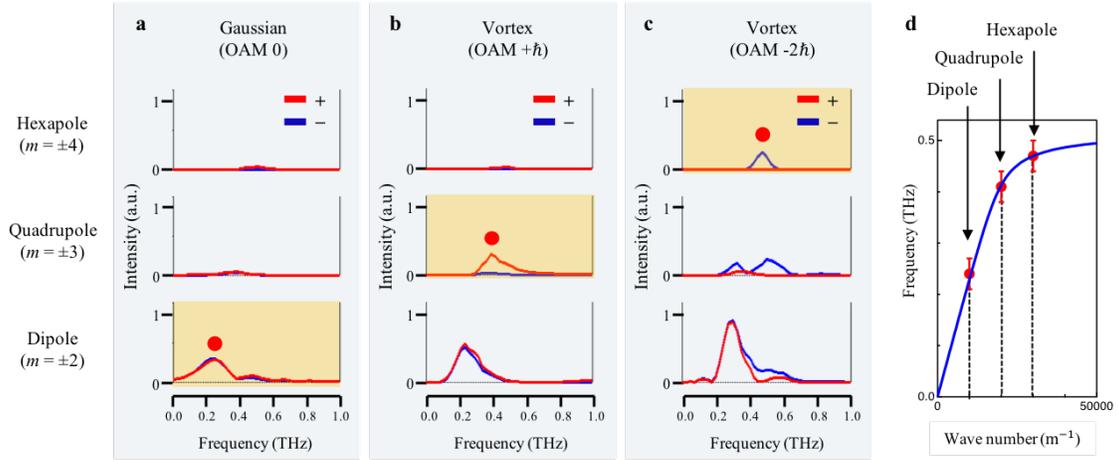

**Figure 3 | Mode decomposition of near field distribution.** Frequency spectra of the dipole [$E(\pm 2, f)$], quadrupole [$E(\pm 3, f)$], and hexapole [$E(\pm 4, f)$] modes excited in the sample illuminated by **a.** Gaussian beam, **b.** vortex beam (+$\hbar$), and **c.** vortex beam (-2$\hbar$). The red (positive) and blue (negative) traces show the traveling wave modes in the clockwise and counter-clockwise direction, respectively. **d.** Dispersion relation of the spoof LSP. The red dots represent the resonance frequencies determined in a-c. The blue curve is a theoretical fitting.

**Selection rules**

These results allow us to deduce the selection rules for the excitation of multipole spoof LSPs. Table 1 summarizes the z component of the angular momentum of the excitation beams (upper part) and the observed LSPs (lower part). Let us first revisit the excitation of the dipole mode by the uniform electric field of the Gaussian beam in terms of the angular momentum. The linearly polarized incident light is a superposition of two circularly polarized waves with the SAM of $+\hbar$ and $-\hbar$. Due to the zero OAM, the total angular momentum (TAM) of the incident Gaussian beam is equal to the SAM. The excited dipole mode, which mimics the dipole LSP with a traveling surface charge distribution $\rho(\varphi,t) \propto \cos(\pm\varphi - 2\pi f t)$, also possesses the $z$-component of the angular momentum of $\pm\hbar$ (see Methods for the relation between the charge and electric field distribution). This gives us an alternative way to interpret the selective excitation of the dipole mode by the Gaussian beam, i.e., transfer of the TAM (= SAM) of light to the electronic system. Similar arguments hold true for the vortex beam excitations. The TAM of the linearly polarized vortex beam (OAM $+\hbar$) is $+2\hbar$ and 0. The clockwise quadrupole mode observed in the experiment also has the $z$-component of the angular momentum of $+2\hbar$, which suggests the transfer of the TAM. Note that no LSP mode is excited by the vortex beam with the TAM of 0 (ref. 11). The TAM of the linearly polarized vortex beam (OAM $-2\hbar$) is $-3\hbar$ and $-\hbar$. The selective excitation of the counter-clockwise hexapole mode with the $z$-component of the angular momentum of $-3\hbar$ obeys the same rule. This rule also predicts that the field component with the TAM of $-\hbar$ at 0.5 THz should excite counter-clockwise dipole mode. This is also observed in Fig. 3c (bottom) where the higher frequency tail of the counter-clockwise dipole mode (blue) is stronger than the other (red). These experimental observations strongly support the selection rules governed by the conservation of the TAM.

|  | Gaussian beam | Vortex beam | | |
| --- | --- | --- | --- | --- |
| Orbital angular momentum | 0 | $+\hbar$ | | $-2\hbar$ |
| Spin angular momentum | $+\hbar$ & $-\hbar$ | | | |
| Total angular momentum | $+\hbar$ & $-\hbar$ | $+2\hbar$ | 0 | $-3\hbar$ | $-\hbar$ |
| ⬇ | ⬇ | ⬇ | | ⬇ | |
| Observed spoof LSP (angular momentum) | Dipole ($+\hbar$ & $-\hbar$) | Quadrupole ($+2\hbar$) | | Hexapole ($-3\hbar$) | Dipole ($-\hbar$) |

**Table 1 | Selection rules for exciting multipolar spoof LSPs.** The orbital (1st row), spin (2nd row), and total angular momentums (3rd row) of the excitation beams are summarized in the upper part. The last row below the downward arrows shows the angular momentums of the spoof LSPs observed in the experiment. Spoof LSP modes having the same angular momentum with the total angular momentum of the excitation light are excited.

**Numerical simulations**

The selection rules inferred by the experiments are exactly the same as the ones predicted for real LSPs by numerical simulations[11]. This tells us that the spoof LSP mimics not only the dispersion relation of the LSP, but also the selection rules for excitation. We numerically confirmed this point by performing similar electromagnetic field analyses as ref. 11 for spoof LSPs (see Methods). Fig. 4a shows an electric field distribution around the sample excited by a linearly polarized Gaussian beam at 0.24 THz. The dipolar standing wave (four nodes) is seen around the outer circle, which reproduces the experimental result (Fig. 4b) very well. Similar field patterns are obtained in simulations performed with different excitation frequencies and the integrated electric field intensities along the outer circle are collected to draw an intensity spectrum (Fig. 4c). This shows the dipolar spoof LSP is resonantly excited at 0.24 THz as observed in the experiment (Fig. 4d). Similar simulations for the vortex beam (OAM $+\hbar$) excitation exhibit the characteristic field distribution (six zero-crossing points) unique to the clockwise quadrupole mode (Fig. 4e), which is very similar to the one in the experimental result (Fig. 4f). The intensity spectrum (Fig. 4g) reveals a very similar resonance frequency (0.39 THz) found from the experiment (Fig. 4h). Similarly, the selective excitation of the counter-clockwise hexapole mode (eight zero-crossing points) is clearly seen in the vortex beam (OAM $-2\hbar$) excitation at 0.52 THz (Fig. 4i). Note the purer hexapolar field pattern than the experimental result (Fig. 4j). The intensity spectrum (Fig. 4k) shows the resonance frequency of the hexapole mode similar to the one revealed by the experimental data (Fig. 4l). At lower excitation frequencies, we start to see the dipolar mode around 0.21 THz, which supports the selection rules described in Table 1. These simulations clearly demonstrate that the same selection rules apply for exciting the spoof LSPs.

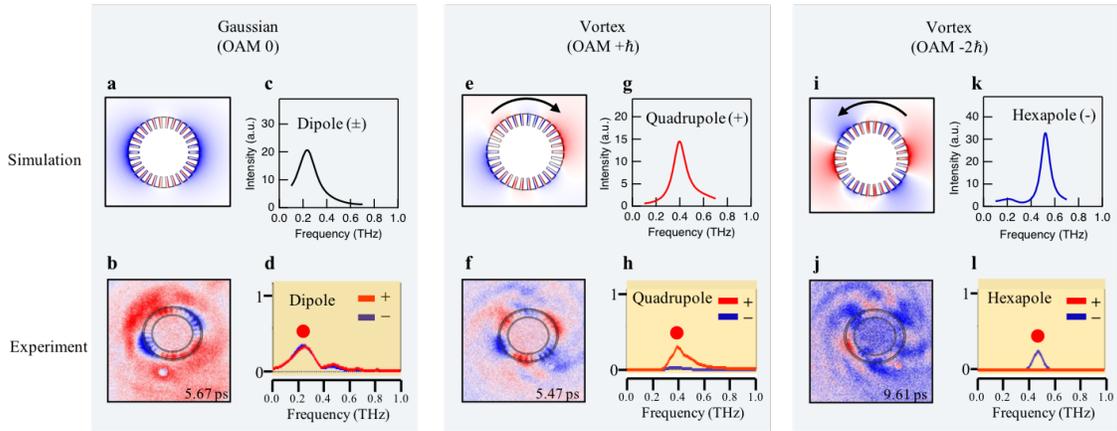

**figure 4 | Numerical simulations showing the selective excitation of multipole spoof LSPs. a.** Simulated electric field distribution around the sample excited by the linearly polarized Gaussian beam (0.24 THz). The figure shows the linear polarization component along the polarization of the excitation beam to simulate the experimental condition. **b.** One experimental snapshot from Fig. 2a is shown for comparison. **c.** Intensity spectrum showing the resonance frequency of the dipole mode at 0.24 THz. **d.** The spectrum of the dipole mode from Fig. 3a is shown for comparison. **e-h.** Similar set of data for the linearly polarized vortex beam (+$\hbar$) excitation. The field pattern (**e**) rotates in the direction shown by the black arrow (clockwise). Note that due to the pure vortex beam excitation, the dipolar mode at 0.24 THz is not excited at all in the spectrum (**g**). **i-l.** Similar set of data for the linearly polarized vortex beam (-2$\hbar$) excitation. The field pattern (**i**) rotates in the direction shown by the black arrow (counter-clockwise).

**Conclusions**

We succeeded in selectively exciting dipole-forbidden eigen modes of spoof LSPs, multipolar modes with high OAM, using the phase vortex beam. The efficient TAM transfer from light to spoof LSPs are achieved. This is due to the negligible electron scattering in metals in the THz frequency region that assures the coherent collective motion of electrons over the entire sample with a similar size to the excitation beam. Theory predicts similar TAM transfer to a variety of elementary excitations in solids such as single electronic states[24], phonons[25], and skyrmions[26]. For the experimental realization of these excitations with subwavelength wavefunctions, focusing techniques beyond the diffraction limit should play a crucial role[27,28].

**References**

[1] Fox, M. *Optical Properties of Solids* (Oxford University Press, 2010).

[2] Hao, F., Larsson, E. M., Ali, T. A., Sutherland, D. S. & Nordlander, P. Shedding light on dark plasmons in gold nanorings. *Chem. Phys. Lett.* **458**, 262–266 (2008).

[3] Takase, M. *et al.* Selection-rule breakdown in plasmon-induced electronic excitation of an isolated single-walled carbon nanotube. *Nat. Photon.* **7,** 550–554 (2013).

[4] Boyd, R. *Nonlinear Optics* (Academic Press, 2008).

[5] Allen, L., Beijersbergen, M. W., Spreeuw, R. J. C., & Woerdman, J. P. Orbital angular momentum of light and the transformation of Laguerre-Gaussian laser modes. *Phys. Rev. A* **45,** 8185–8189 (1992).

[6] Van Enk, S. J. & Nienhuis, G. Commutation rules and eigenvalues of spin and orbital angular momentum of radiation fields. *J. Mod. Opt.* **41**, 963–977 (1994).

[7] Schmiegelow, C. T. & Schmidt-Kaler, F. Light with orbital angular momentum interacting with trapped ions. *Eur. Phys. J. D* **66,** 1–9 (2012).

[8] Schmiegelow, C. T. *et al.* Transfer of optical orbital angular momentum to a bound electron. *Nat. Commun* **7**, 12998 (2016).

[9] Alexandrescu, A., Cojoc, D. & Di Fabrizio, E. Mechanism of angular momentum exchange between molecules and Laguerre-Gaussian beams. *Phys. Rev. Lett.* **96**, 243001

(2006).

[10] Shigematsu, K., Yamane, K., Morita, R. & Toda, Y. Coherent dynamics of exciton orbital angular momentum transferred by optical vortex pulses. *Phys. Rev. B* **93,** 045205 (2016).

[11] Sakai, K., Nomura, K., Yamamoto, T. & Sasaki, K. Excitation of Multipole Plasmons by Optical Vortex Beams. *Sci. Rep.* **5**, 8431 (2015).

[12] Pors, A., Moreno, E., Moreno, L. M., Pendry, J. B. & Vidal, F. J. G. Localized spoof plasmons arise while texturing closed surfaces. *Phys. Rev. Lett.* **108,** 223905 (2012).

[13] Shen, X. & Cui, T. J. Ultrathin plasmonic metamaterial for spoof localized surface plasmons. *Laser Photonics Rev.* **8,** 137 (2014).

[14] Chen, L., Wei, Y., Zang, X., Zhu, Y. & Zhuang, S. Excitation of dark multipolar plasmonic resonances at terahertz frequencies. *Sci. Rep.* **6**, 22027 (2016).

[15] Blanchard, F. *et al.* Real-time terahertz near-field microscope. *Opt. Express* **19,** 8277–8284 (2011).

[16] Blanchard, F., Doi, A., Tanaka, T. & Tanaka, K. Real-time, subwavelength terahertz imaging. *Annu. Rev. Mater. Res.* **43,** 237–259 (2013).

[17] Blanchard, F. & Tanaka, K. Improving time and space resolution in electro-optic sampling for near-field terahertz imaging. *Opt. Lett.* **41,** 4645–4648 (2016).

[18] Beijersbergen, M. W., Coerwinkel, R. P. C., Kristensen, M. & Woerdman, J. P. Helical-wavefront laser beams produced with a spiral phaseplate. *Opt. Commun.* **112,** 321–327 (1994).

[19] Turnbull, G. A., Robertson, D. A., Smith, G. M., Allen, L. & Padgett, M. J. The generation of free-space Laguerre-Gaussian modes at millimetre-wave frequencies by use of a spiral phaseplate. *Opt. Commun.* **127,** 183–188 (1996).

[20] Miyamoto, K., Suizu, K., Akiba, T. & Omatsu, T. Direct observation of the topological charge of a terahertz vortex beam generated by a Tsurupica spiral phase plate. *Appl. Phys. Lett.* **104,** 261104 (2014).

[21] Miyamoto, K. *et al.* Highly intense monocycle terahertz vortex generation by utilizing a Tsurupica spiral phase plate. *Sci. Rep.* **6,** 38880 (2016).

[22] Vidal, F. J. G., Moreno, L. M. & Pendry, J. B. Surfaces with holes in them: new plasmonic metamaterials. *J. Opt. A: Pure Appl. Opt.* **7,** S97 (2005).

[23] Rusina, A., Durach, M. & Stockman, M. I. Theory of spoof plasmons in real metals. *Appl. Phys. A* **2,** 375–378 (2010).

[24] Quinteiro, G. F. & Tamborenea, P. I. Electronic transitions in disk-shaped quantum dots induced by twisted light, *Phys. Rev. B* **79,** 155450 (2009).

[25] Zhang, L. & Niu, Q. Chiral phonons at high-symmetry points in monolayer hexagonal lattices. *Phys. Rev. Lett.* **115,** 115502 (2015).

[26] Fujita, H. & Sato, M. Ultrafast generation of skyrmionic defects with vortex beams: Printing laser profiles on magnets. *Phys. Rev. B* **95,** 054421 (2017).

[27] Arikawa, T., Morimoto, S. & Tanaka, K. Focusing light with orbital angular momentum by circular array antenna. *Opt. Express* **25,** 13728-13735 (2017).

[28] Sakai, K., Yamamoto, T. & Sasaki, K. Nanofocusing of structured light for quadrupolar light-matter interactions. *Sci. Rep.* **8,** 7746 (2018).

**Acknowledgements**

This work was supported by KAKENHI (26247052, 17H06124) from Japan Society for the Promotion of Science (JSPS) and JST CREST. This work was supported in part by Kyoto University Nano Technology Hub in "Nanotechnology Platform Project" sponsored by the Ministry of Education, Culture, Sports, Science and Technology (MEXT), Japan.

**Author Contributions**

T.H, S.M performed the measurements and analyzed the data under the supervision and guidance of T.A, F.B. and K.T. S.T and T.T helped initial measurements and data analysis. K. Sakai, H.K. and K. Sasaki performed numerical simulations. T.A. and K.T. prepared the manuscript. All authors discussed the results and contributed to the manuscript.

**Methods**

**Spiral phase plate.** We used two SPPs made of ZEONEX (cyclo olefin polymer, refractive index 1.52) with different step heights. SPP-1 is nominally designed to convert the Gaussian beam into the vortex beam ($\pm\hbar$) at 0.45 THz. The step height is 1.29 mm. SPP-2 has a step of 2.18 mm, which generates the vortex beam ($\pm 2\hbar$) at 0.53 THz. The steps are discretized to 16 small steps. The sign of the OAM can be inverted by reversing the SPP in the beam path.

**Azimuthal angle dependence of the local electric field around multipolar LSPs.** As schematically shown in Fig. 2b,d,f, the surface charge distributions of LSPs along the circumference (radius $r = r_0$) can be expressed as follows,

$$\rho_l(\varphi) \propto \cos(l\varphi) \quad (1)$$

Here $l = 1, 2, ...$ corresponds to the dipole, quadrupole and higher order LSPs, respectively. We ignore the time-dependence (rotation of the charge distribution) at this moment and calculate the static electric field, $\vec{E}_l(r,\varphi)$. The scalar potential $\Phi(r,\varphi)$ can be obtained by solving the Laplace equation (at $r \neq r_0$) in cylindrical coordinates ($r, \varphi, z$). Due to the cosine-type charge distribution, the scalar potential is an even function about $\varphi$. When the scalar potential is independent of $z$, the general solution is,

$$\Phi(r,\varphi) = A_0 + B_0 \log r + \sum_{l=1}^{\infty}(A'_l r^l + A_l r^{-l}) \cos(l\varphi) \quad (2)$$

where $A_0$, $B_0$, $A_l$, and $A'_l$ are coefficients determined by boundary conditions. In order to avoid the divergence of the electric field in the limit of $r \to 0$ and $r \to \infty$, the electric field distributions have to take the following forms,

$$\vec{E}(r,\varphi) = \begin{cases} \sum_{l=1}^{\infty} l A'_l r^{l-1}\left[-\cos(l\varphi)\,\vec{e}_r + \sin(l\varphi)\,\vec{e}_\varphi\right] & (r \leq r_0) \\ \sum_{l=1}^{\infty} l A_l r^{-l-1}\left[\cos(l\varphi)\,\vec{e}_r + \sin(l\varphi)\,\vec{e}_\varphi\right] - \frac{B_0}{r}\vec{e}_r & (r \geq r_0) \end{cases}$$

(3)

From the boundary condition, $[\vec{E}(r \geq r_0, \varphi) - \vec{E}(r \leq r_0, \varphi)] \cdot \vec{e}_r = \rho_l(\varphi)$ at $r = r_0$, we obtain the following result for the static electric field around $\rho_l(\varphi)$,

$$\vec{E}_l(r, \varphi) = lA_l r^{-l-1}[\cos(l\varphi)\vec{e}_r + \sin(l\varphi)\vec{e}_\varphi] \qquad (r \geq r_0) \quad (4)$$

By taking the projection onto the detection polarization axis $\vec{e}_0 = \vec{e}_r\cos(\varphi) - \vec{e}_\varphi\sin(\varphi)$, the azimuthal angle dependence of the electric field detected in the experiment is,

$$\vec{E}_l(\varphi) \cdot \vec{e}_0 \propto \cos((l+1)\varphi) \tag{5}$$

When $\rho_l(\varphi)$ rotates as $\rho_l(\varphi,t) = \cos(l\varphi \mp 2\pi ft) = \cos(\pm l\varphi - 2\pi ft)$, additional electric field components such as radiation field and induced field are generated. However, in the near field regime, the following quasi-static field is the dominant one,

$$E_l(\varphi, t) \propto \cos(\pm(l+1)\varphi - 2\pi ft) \tag{6}$$

By replacing $\pm(l+1)$ by $m = \pm 2$ (dipole), $\pm 3$ (quadrupole), and so on, we have

$$E_m(\varphi, t) \propto \cos(m\varphi - 2\pi ft) \tag{7}$$

As mentioned in the main text, this shows different LSP modes correspond one-to-one to cosine functions which are orthogonal to each other. Similar calculation for $l = 0$, i.e., a uniform surface charge $\rho_0(\varphi) \propto$ Const. leads to the following form,

$$E_1(\varphi, t) \propto \cos(\varphi)\cos(2\pi ft) \tag{8}$$

These functions $E_m(\varphi, t)$ ($m = \pm 1, \pm 2, \pm 3, ...$) plus $E_0(\varphi, t) = $ Const. form a complete set for symmetric field distributions about $\varphi = 0$ with the period of $2\pi$.

**Dispersion relation of spoof LSPs.**

The dispersion relation of the spoof surface plasmons that propagate on the periodically corrugated metal surfaces is given as follows[22,23],

$$k = \frac{2\pi f n_{out}}{c} \sqrt{1 + \left(\frac{a}{d}\frac{n_{out}}{n_g}\right)^2 \tan^2\left(\frac{2\pi f n_g}{c}(R-r)\right)} \qquad (9)$$

Here, $k$ is the wavevector along the surface, and $c$ is the speed of light. The asymptotic (cutoff) frequency in the limit of $k \to \infty$ is $c/4n_g(R-r)$. By folding the surface to make a corrugated metallic disk, $k$ becomes the wavevector along the circumference, which is quantized due to the periodic boundary condition. We used eq. (9) as a fitting function in Fig. 3d. The fitting parameters determined by the method of least squares are $n_g = 4.8$ and $n_{out} = 2.1$. These numbers represent the effective refractive indices for the electromagnetic waves inside the groove and outside the disk. These should be expressed by combinations of the refractive indices of air ($n = 1.00$), silicon ($n = 3.41$), glass ($n = 1.96$), high-reflection coating materials ($n \sim 2$) and LN ($n = 5.11$) with appropriate weightings depending on the spatial extent of the electric field (see Fig. 1c for the structure around the metallic disk). The high effective index inside the groove ($n_g = 4.8$) indicates that the electromagnetic mode is strongly localized around the LN ($n = 5.11$). The effective index outside the disk ($n_{out} = 2.1$), which is close to the index of the glass substrate ($n = 1.96$) suggests that the localization is weaker.

**Numerical simulation.**

We numerically calculated the electromagnetic field in our system using the finite-element method in the commercial software COMSOL Multiphysics with an RF module. We performed three-dimensional harmonic propagation calculation with perfectly matched layers (PMLs) on the side and lower boundaries, plus the scattering boundary condition on the top surface. At the top surface, we excited the THz incident beam and made it propagate in the z direction, normal to the EO crystal. The perfect electric conductor was used at the surface of the corrugated disk to simulate the spoof LSP. To ensure the good match in the resonance frequency between the experiment and the simulation, an air layer of 1 μm thickness was introduced above the corrugated disk.

Supplementary information

# Selective excitation of multipolar spoof plasmons using orbital angular momentum of light

Takashi Arikawa[1†], Tomoki Hiraoka[1†], Shohei Morimoto[1†],
François Blanchard[2], Shuntaro Tani[3], Tomoko Tanaka[3],
Kyosuke Sakai[4], Hiroki Kitajima[4], Keiji Sasaki[4] and Koichiro Tanaka[1,3] *

[1] Kyoto University, Department of Physics, Kyoto, 606-8502, Japan
[2] École de technologie supérieure (ÉTS), Department of Electrical Engineering, Montréal, Québec, H3C 1K3, Canada
[3] Kyoto University, Institute for Integrated Cell-Material Sciences (iCeMS), Kyoto, 606- 8501, Japan
[4] Research Institute for Electronic Science, Hokkaido University, Hokkaido, 001-0020, Japan

† These authors contributed equally to this work
* email: kochan@scphys.kyoto-u.ac.jp

**Table of contents**

# 1. Experimental setup

Figure S1 schematically shows the experimental setup. We used a mode-locked Ti:Sapphire regenerative amplifier (Solstice, Spectra-Physics) that delivers 100 fs optical pulses (center wavelength 780 nm) with 3.2 mJ pulse energy at 1 kHz repetition rate. Linearly polarized coherent THz pulses with Gaussian beam profile were generated by optical rectification in a LiNbO3 (LN) crystal using tilted-pulse-front excitation scheme [1,2]. We used 1-μm-thick x-cut LN as an electro-optic (EO) crystal to detect electric field of THz pulses[3]. The image of the probe beam at the EO crystal was relayed to a 16 bit CMOS camera from PCO (model PCO Edge) with a polarization analyzer unit. The short pass (SP) and long pass (LP) filters were used to enhance the detection sensitivity using the probe spectrum filtering technique[4]. An x-cut LN with the thickness of 2 μm is used to compensate for the birefringence of the EO crystal.

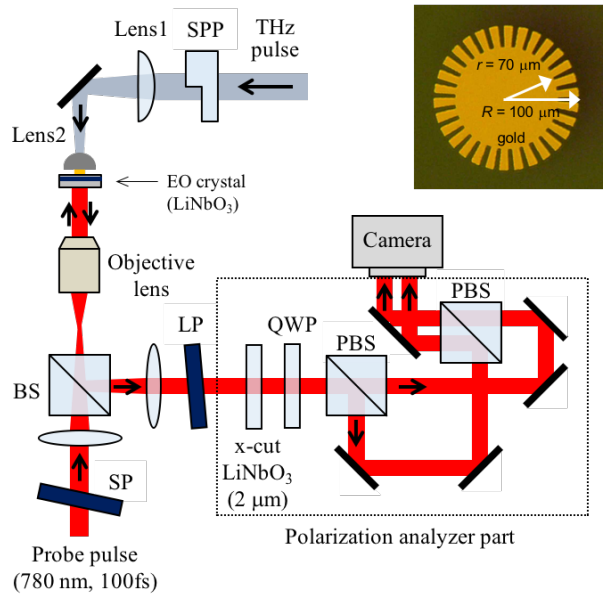

**Figure S1 | Experimental setup.** Optical system for the THz near-field microscope. SPP: spiral phase plate, BS: beam splitter, SP: short pass filter, LP: long pass filter, QWP: quarter wave plate, PBS: polarizing beam splitter. The optical image of the sample used in the main text is shown in the upper right corner.

## 2. OAM purity of the excitation beam

The OAM purity of the excitation beam can be evaluated by the same field pattern analysis performed in the main text. The drawings on the top of Fig. S2a schematically shows the electric field vector ($\vec{E}$, grey arrows) of the linearly polarized Gaussian beam and vortex beams with the OAM of $\pm\hbar$ and $\pm 2\hbar$. These are drawn at a time $t = t_0$ when the electric field at $\varphi = 0$ position is positive and maximum. The azimuthal angle dependence of the electric field along a virtual dashed circle are also plotted [$E(\varphi, t = t_0) = \vec{E} \cdot \vec{e}_0$, $\vec{e}_0$: polarization direction detected in the experiment]. As shown in these graphs, different OAM modes are characterized by different orthogonal functions [$\cos(l\varphi)$, $l = 0$; Gaussian, $l = \pm 1$; vortex ($\pm\hbar$), $l = \pm 2$; vortex ($\pm 2\hbar$)]. This is evident because the electric field of the vortex beams with the OAM of $l\hbar$ have the azimuthal angle dependence of $\exp(il\varphi)$. This enables us to analyze the OAM purity of the excitation beam.

We performed two-dimensional Fourier transform of $E(\varphi, t)$ taken from the excitation beams (Fig 1d,g,j) to obtain frequency spectra $E(l, f)$. The same radius as the outer radius of the corrugated metallic disk used in the main text was used for the radius of the virtual circle. Figure S2b shows the frequency spectra of the Gaussian [$E(0, f)$], $\pm\hbar$ vortex [$E(\pm 1, f)$], and $\pm 2\hbar$ vortex [$E(\pm 2, f)$] components. In the case of no SPP, the spectra show that the beam is very close to the Gaussian beam. With SPP-1, most of the frequency component around $0.55 \pm 0.2$ THz is converted to the vortex beam ($+\hbar$). The spectral shape depends on the spectrum of the original beam before going through the SPP. This explains the center frequency of 0.55 THz although SPP-1 is nominally designed at 0.45 THz. The residual Gaussian component around 0.25 THz is responsible for the dipole mode observed in Fig. 3b. SPP-2 generates vortex beam ($-2\hbar$) around $0.55 \pm 0.1$ THz. As expected, it also generates vortex beam ($-\hbar$) around the half frequency region. This component excites the quadrupole mode observed in Fig. 3c. The vortex beam ($-\hbar$) component around 0.6 THz is presumably due to the imperfection of the SPP.

As shown in the main text and above, $\cos(\pm 2\varphi)$ component contains information on the dipole LSP mode and vortex beam ($\pm 2\hbar$). This is the reason why we eliminate the excitation beam signal in time-domain for the LSP mode identification done in the main text.

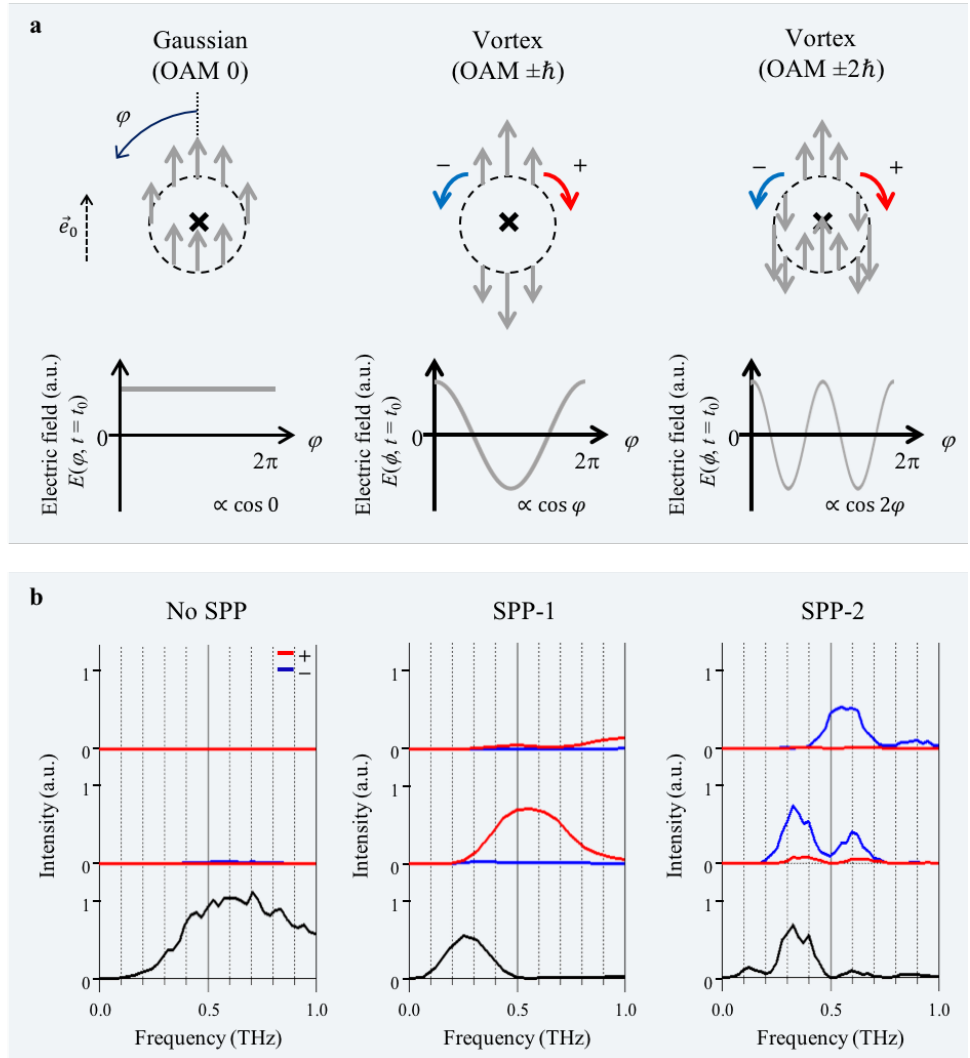

**Figure S2 | Field pattern analysis of the excitation beam. a.** Electric field vector (grey arrows) of the linearly polarized Gaussian beam and vortex beams with the OAM of $\pm\hbar$ and $\pm2\hbar$. The black cross represents the centre position of the beam and virtual dashed circle. The azimuthal angle dependence of the electric field for each beam is shown below. **b.** Frequency spectra of the Gaussian, vortex ($\pm\hbar$), and vortex ($\pm2\hbar$) components obtained from the data shown in Fig. 1d (No SPP), Fig. 1g (SPP-1), and Fig. 1j (SPP-2).

## 3. Additional data
3.1 Excitation of inverse rotation modes

Figure S3a shows frequency spectra of the quadrupole mode excited by the vortex beam with the OAM of $+\hbar$ and $-\hbar$. These data clearly show that we can selectively excite both clockwise and counter-clockwise modes depending on the sign of the OAM. These results also obey the same selection rules discussed in the main text. The similar data set in Fig. S3b for the hexapole mode excited by the vortex beam with the OAM of $+2\hbar$ and $-2\hbar$ also shows the same selectivity governed by the conservation of the TAM.

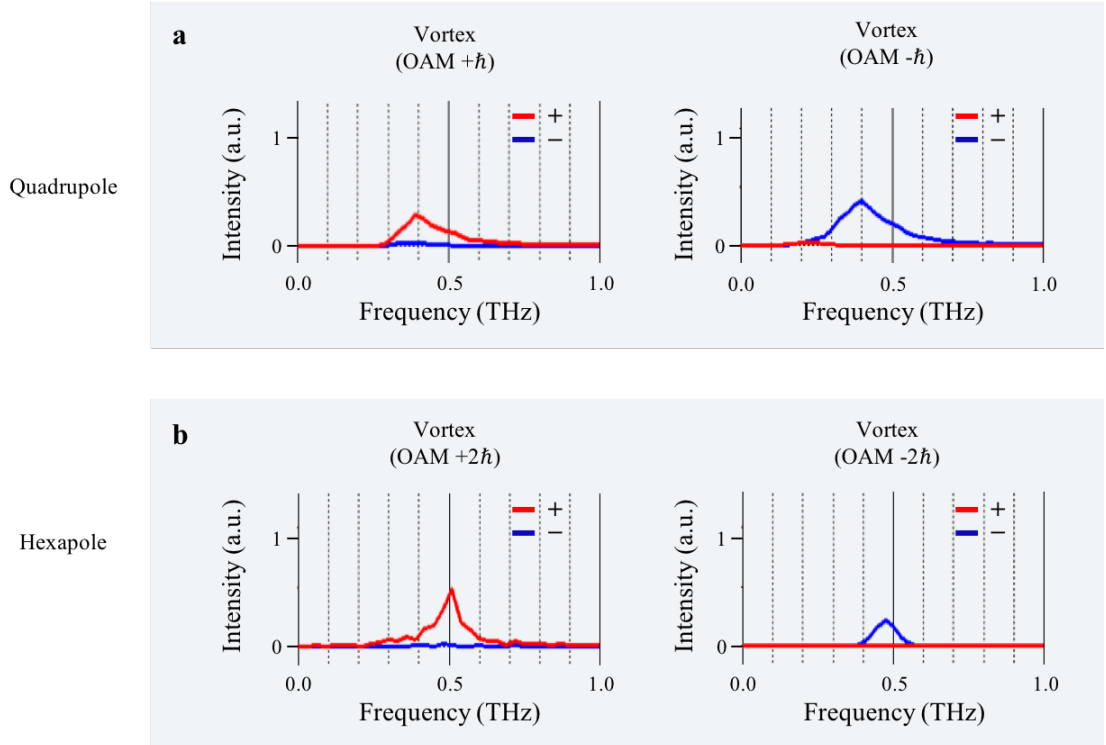

**Figure S3 | Selection of inverse rotation modes. a.** Frequency spectra of the quadrupole mode excited by vortex beams with the OAM of $+\hbar$ and $-\hbar$. The left graph is the same as the one shown in Fig. 3b (middle). **b.** Frequency spectra of the hexapole mode excited by vortex beams with the OAM of $+2\hbar$ (left) and $-2\hbar$ (right). The right graph is the same as the one shown in Fig. 3c (top). All the data were taken with the same sample as the one used in the main text.

3.2 Groove depth dependence

    As described in the Methods section, the cutoff frequency of the spoof LSP is inversely proportional to the groove depth ($R - r$). Accordingly, the resonance frequency of the spoof LSP is expected to increase as the groove depth becomes shallow. To verify this point, we fabricated corrugated metallic disks with differing groove depths (outer radius $R$ = 100 μm is fixed). Figure S4a shows the spectra of the dipole mode observed in three different samples drawn on the left. We can see that the resonance frequency shifts to the higher frequency as the groove becomes shallower. Similar trends are observed for quadrupole and hexapole modes as shown in Fig. S4b,c, respectively.

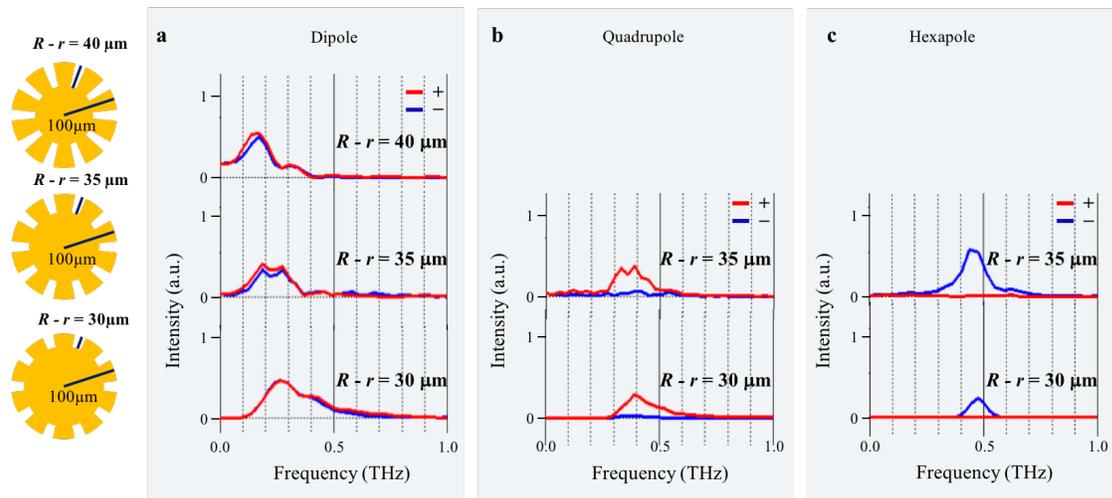

**Figure S4 | Groove depth dependence. a.** Frequency spectra of the dipole mode excited in three different samples. Gaussian beam is used for the excitation. **b.** Frequency spectra of the quadrupole mode excited in two different samples. Vortex beam ($+\hbar$) is used for the excitation. **c.** Frequency spectra of the hexapole mode excited in two different samples. Vortex beam ($-2\hbar$) is used for the excitation.

3.3 Outer radius dependence

    Figure S5 shows how the resonance frequency of the spoof LSP mode changes depending on the outer radius of the metallic disk. We fabricated three samples with different outer radius while maintaining the groove depth constant (35 μm). As shown in Fig. S5a, the resonance frequency of the dipole mode shifts to the higher frequency as the outer radius decreases; shorter circumference leads to a LSP with shorter wavelength. The same trend is observed for the quadrupole mode (Fig. S5b).

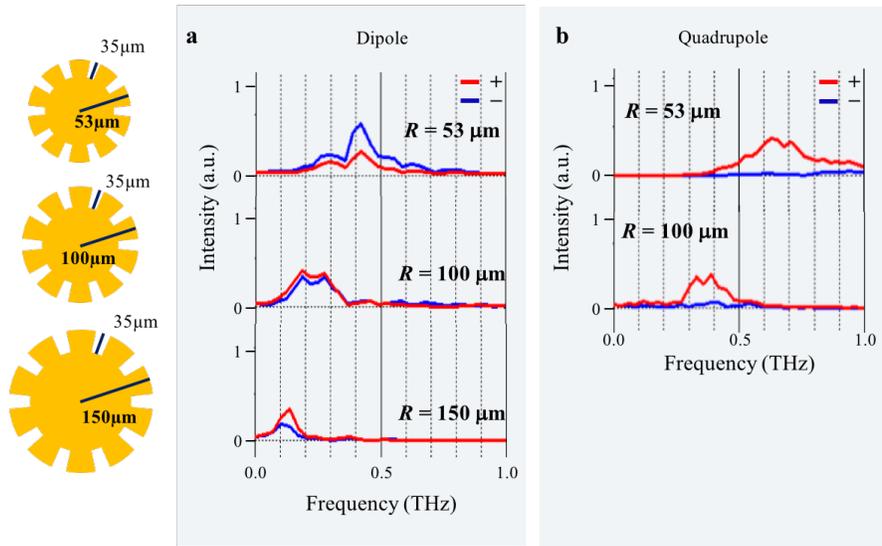

**Figure S5 | Outer radius dependence. a.** Frequency spectra of the dipole mode excited in three different samples. Gaussian beam is used for the excitation. **b.** Frequency spectra of the quadrupole mode excited in two different samples. Vortex beam (+ℏ) is used for the excitation.

# 4. References

[1] Hebling, J., Almási, G., Kozma, I., & Kuhl, J. Velocity matching by pulse front tilting for large area THz-pulse generation. *Opt. Express* **10**, 1161–1166 (2002).

[2] Hirori, H., Doi, A., Blanchard, F., & Tanaka, K. Single-cycle terahertz pulses with amplitudes exceeding 1 MV/cm generated by optical rectification in LiNbO3. *Appl. Phys. Lett.* **98**, 091106 (2011).

[3] Winnewisser, C., Jepsen, P. Uhd, Schall, M., Schyja, V., & Helm, H. Electro-optic detection of THz radiation in $LiTaO_3$, $LiNbO_3$ and ZnTe. *Appl. Phys. Lett.* **70**, 3069 (1997).

[4] Blanchard, F., & Tanaka, K. Improving time and space resolution in electro-optic sampling for near-field terahertz imaging. *Opt. Lett.* **41**, 4645–4648 (2016).